\documentclass{article}

\usepackage{arxiv}

\usepackage[utf8]{inputenc} 
\usepackage[T1]{fontenc}    
\usepackage{hyperref}       
\usepackage{url}            
\usepackage{booktabs}       
\usepackage{amsfonts}       
\usepackage{nicefrac}       
\usepackage{microtype}      
\usepackage{lipsum}
\usepackage{graphicx}

\usepackage{amsmath}  
\usepackage{multirow}
\usepackage{tabularx}
\usepackage{amssymb}
\usepackage{booktabs} 
\newcommand{\hide}[1]{\iffalse #1 \fi}

\usepackage{latexsym}
\usepackage{xcolor}
\usepackage{ulem} 

\graphicspath{ {./images/} }

\title{PGDiffSeg: Prior-Guided Denoising Diffusion Model with Parameter-Shared Attention for Breast Cancer Segmentation}

\author{
 Feiyan Feng \\
  School of Information \\ Science and Engineering, \\Shandong Normal University, \\Jinan 250358, China \\
  \texttt{1720696671@qq.com} \\
   \And
 Tianyu Liu \\
  School of Information \\ Science and Engineering, \\Shandong Normal University, \\Jinan 250358, China \\
  \texttt{lty.tianyu@qq.com} \\
  \And
 Hong Wang* \\
  School of Information \\ Science and Engineering, \\Shandong Normal University, \\Jinan 250358, China \\
  \texttt{111052@sdnu.edu.cn} \\
     \And
 Jun Zhao \\
  School of Information \\ Science and Engineering, \\Shandong Normal University, \\Jinan 250358, China \\
  \texttt{zhaojun@sdnu.edu.cn} \\
  \And
 Wei Li \\
  School of Information \\ Science and Engineering, \\Shandong Normal University, \\Jinan 250358, China \\
  \texttt{1966067505@qq.com} \\
  \And
 Yanshen Sun* \\
  Department of Computer Science, \\ Virginia Tech,Virginia, \\ 24061, USA \\
  \texttt{yansh93@vt.edu} \\
}

\begin{document}
\maketitle
\begin{abstract}
Early detection through imaging and accurate diagnosis is crucial in mitigating the high mortality rate associated with breast cancer. However, locating tumors from low-resolution and high-noise medical images is extremely challenging. Therefore, this paper proposes a novel PGDiffSeg (\underline{P}rior-\underline{G}uided \underline{Diff}usion Denoising Model with Parameter-Shared Attention) that applies diffusion denoising methods to breast cancer medical image segmentation, accurately recovering the affected areas from Gaussian noise. Firstly, we design a parallel pipeline for noise processing and semantic information processing and propose a parameter-shared attention module (PSA) in multi-layer that seamlessly integrates these two pipelines. This integration empowers PGDiffSeg to incorporate semantic details at multiple levels during the denoising process, producing highly accurate segmentation maps. Secondly, we introduce a guided strategy that leverages prior knowledge to simulate the decision-making process of medical professionals, thereby enhancing the model's ability to locate tumor positions precisely. Finally, we provide the first-ever discussion on the interpretability of the generative diffusion model in the context of breast cancer segmentation. Extensive experiments have demonstrated the superiority of our model over the current state-of-the-art approaches, confirming its effectiveness as a flexible diffusion denoising method suitable for medical image research. Our code will be publicly available later.
\end{abstract}


\begin{figure*}[!t]
\centering
\includegraphics[scale=.5]{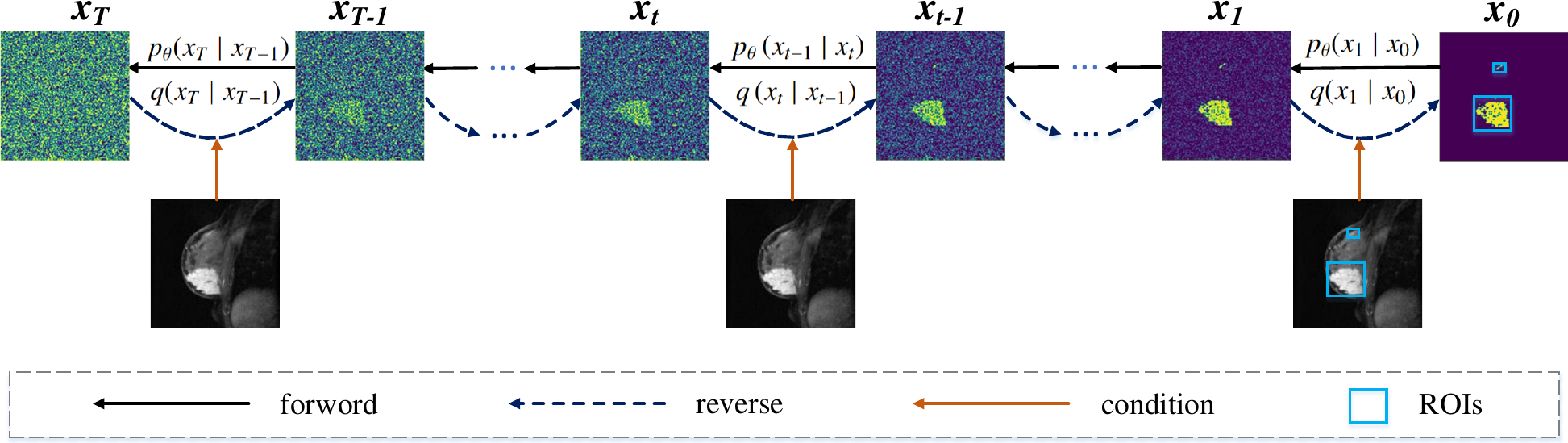}
\caption{The process of PGDiffSeg. It is visualized using a viridis color mapping, where the color ranges from deep purple (representing 0) to bright yellow (representing 1). This process includes both the forward process (noising) and the reverse process (denoising). At each step, a certain amount of noise is added until the image becomes a Gaussian distribution. The model learns denoising schemes using images as conditions and generates regions of interest (ROIs).}
\label{fig1:process_of_diffusion}
\end{figure*}

\section{Introduction}
\label{sec1}
According to the 2023 cancer statistics report \cite{https://doi.org/10.3322/caac.21763}, breast cancer ranks second in mortality rate among female cancers, just behind lung cancer. Nevertheless, extensive research has shown that early screening plays a pivotal role in enhancing the 5-year relative survival rate because initial morphology and location of tumors can be detected through medical imaging \cite{doi:10.1126/science.aay9040}. This early detection is instrumental in achieving more favorable treatment outcomes \cite{CROSWELL2010202, https://doi.org/10.1002/ijc.29634}. Furthermore, advancements in treatment protocols, including targeted therapies and adjuvant chemotherapies, have contributed to the rapid increase in the survival rate of breast cancer patients \cite{SASAKI2015e186}. However, despite these positive developments, the latest research reveals a concerning phenomenon: the exponential increase in medical images, including mammography, magnetic resonance imaging (MRI) sequences, and breast ultrasound images, has significantly extended the time required for radiologists to interpret and analyze these images, which overwhelms radiologists and leads to a high misdiagnosis rate \cite{5656974}. Accurately pinpointing tumors from low-resolution and noisy breast images is challenging \cite{10130343}, requiring substantial effort and expertise for precise tumor segmentation in medical images \cite{9154595}.

In recent years, deep learning-based methods, such as deep convolutional neural networks (DCNNs), have emerged as powerful tools for segmenting suspected regions and extracting features from segmented breast cancer ROIs \cite{9356353}. Integrating computer-aided diagnosis (CAD) strategies based on DCNNs can assist radiologists in making informed decisions, ultimately reducing unnecessary biopsies and alleviating patients' discomfort caused by unnecessary invasive procedures.

The U-shape architecture has shown significant progress in breast cancer segmentation by effectively integrating low-level and high-level information through skip connections. Inspired by its success, many breast cancer segmentation networks based on UNet have sprung up \cite{9669083, 8379359, 10.1007/978-3-030-66415-2_27}.
However, current CAD methods face several challenges. Firstly, they tend to rely excessively on pre-processing phases, such as image denoising and enhancement, resulting in image distortion and the loss of crucial primitive features. In the context of breast cancer segmentation, the utilization of high-resolution images that capture various cancer-related features has the potential to enhance the discrimination ability of the model \cite{9444895}. Moreover, end-to-end segmentation methods lack interpretability due to their "black box" feature \cite{LIU2023105145}. Radiologists may question the authenticity of the segmentation results since the decision-making process needs more guidance or expert knowledge. Furthermore, specific hybrid methods focus on exploring additional information, such as global and local features, that are relevant to the object areas. These methods extract cancer-related features from breast images and directly sum different representations. However, these approaches can widen the semantic gap and neglect essential medical commonsense knowledge.


To address these challenges, we introduce a novel GFDiffSeg model, \textbf{P}rior-\textbf{G}uided \textbf{Diff}usion Denoising Model with Parameter-Shared Attention, for breast cancer segmentation. Unlike the classical Denoising Diffusion Probabilistic Model (DDPM) \cite{NEURIPS2020_4c5bcfec} \hide{[12]}, GFDiffSeg goes beyond iteratively transforming a noise-corrupted input into a clean sample using the diffusion process and probabilistic modeling. It also leverages images as conditions to learn the denoising process for corresponding segmentation masks. The process of noising and denoising in GFDiffSeg is depicted in Fig. \ref{fig1:process_of_diffusion}. By incorporating this principle, GFDiffSeg can directly obtain segmentation results from the raw images while providing interpretability. Furthermore, we propose a parameter-shared attention (PSA) module to effectively fuse noise and image features, bridging the semantic gap between different representations. Additionally, research has demonstrated that incorporating prior knowledge into networks can provide benefits \cite{9785606, 9455423}, so we suggest adopting a prior knowledge-guided strategy to facilitate rapid focus on regions of interest. These advancements in the GFDiffSeg model contribute to improved segmentation performance and efficiency in breast cancer analysis. Our contributions can be summarized as follows:

\begin{itemize}
 
    \item PGDiffSeg represents a pioneering approach that enhances diffusion denoising techniques to achieve accurate breast cancer medical image segmentation by effectively eliminating Gaussian noise. Notably, PGDiffSeg holds the distinction of being the first diffusion generative model in the field of breast cancer segmentation, equipped with valuable interpretability capabilities.
    \item We introduce a parallel pipeline for separate processing of image and noise-added label information. Furthermore, we design a PSA module that enhances the model's focus on lesion areas in both pipelines to address the fusion of noise features and semantic information. This integration empowers PGDiffSeg to incorporate semantic details at multiple levels during denoising. 
    
    \item We incorporate prior knowledge by utilizing context encoding, allowing our model to simulate doctors' decision-making process. This integration of prior knowledge successfully reduces the segmentation difficulty for our model. This module operates akin to the expertise of physicians, guiding the denoising process and facilitating quicker focus on lesion areas.
    
    \item We are the first to present the interpretability of diffusion models and analyze the attention transfer characteristics in image segmentation tasks. In contrast to end-to-end models, we visualize intermediate results and attention positions at each iteration, providing a clear view of the model's learning process and enhancing trust among medical professionals\hide{(see Fig. \ref{sampling-process})}. 
    
    \item Our method is extensively evaluated and consistently outperforms or achieves comparable results to the current state-of-the-art models in breast cancer image segmentation. Moreover, experiments reveal our model's excellent transferability, allowing it to be seamlessly applied to diverse modalities such as MRI or ultrasound datasets.
\end{itemize}

We organize the rest of this paper as followings. The related works are reviewed in Section \ref{sec2}. The proposed model and its detailed workflow are presented in Section \ref{sec3}. The experiments, and analysis are demonstrated in Sections \ref{sec4} and \ref{sec6}. Ultimately, we summarize our work and provide future perspectives in Section \ref{sec7}.

\section{Related Work}
\label{sec2}
\subsection{Diffusion model in medical image processing tasks}

Due to its remarkable ability to generate realistic images, diffusion model has captured considerable attention in computer vision \cite{gong2022pet, CHUNG2022102479, amit2022segdiff, zimmermann2021scorebased, 10.1007/978-3-031-16431-6_51, Moghadam_2023_WACV}. There are three generic diffusion modeling frameworks, each based on denoising diffusion probabilistic models, noise-conditioned score networks, and stochastic differential equations \cite{10081412}. When performing segmentation tasks, diffusion models add noise to the segmentation mask and use the image as an additional input to denoise the noisy mask. However, existing segmentation methods are inadequate in maximizing the use of image information to improve segmentation performance. They either concatenate the image and \( x_t \) directly \cite{pmlr-v172-wolleb22a}, add the encoded image features and encoded \( x_t \) \cite{amit2022segdiff}, embed the \( x_t \) features into the image features at each downsampling step \cite{wu2024medsegdiff}, or use cross-attention to fuse the image and \( x_t \) before downsampling \cite{chowdary2023diffusion}. Therefore, this study innovatively designs a fusion strategy where image information and noisy labels mutually reinforce each other,  facilitating the generation of enhanced and more effective feature representations.

\subsection{Self-attention mechanism in medical image segmentation}

The self-attention mechanism has gained significant popularity in medical image analysis due to its ability to capture long-range dependencies and global features. Consequently, several pure transformer models have been developed for medical image segmentation tasks \cite{10.1007/978-3-030-87193-2_8, 10.1007/978-3-031-25066-8_9}. However, these pure transformer designs often neglect the extraction of local information. Recent research has explored hybrid architectures that combine self-attention with CNNs for more effective segmentation \cite{10.1007/978-3-030-87193-2_4, 10.1007/978-3-030-87193-2_2, 9785614, Liu_2021_ICCV, 9706678}. Nonetheless, these studies primarily focus on extracting multi-scale information without effectively addressing the interaction of such information. In contrast, our proposed PSA module emphasizes the fusion of two distinct types of information, thereby improving segmentation performance by bridging the semantic gap.

\section{PGDiffSeg model}
\label{sec3}

\begin{figure*}[htbp]
\centering
\includegraphics[scale=.55]{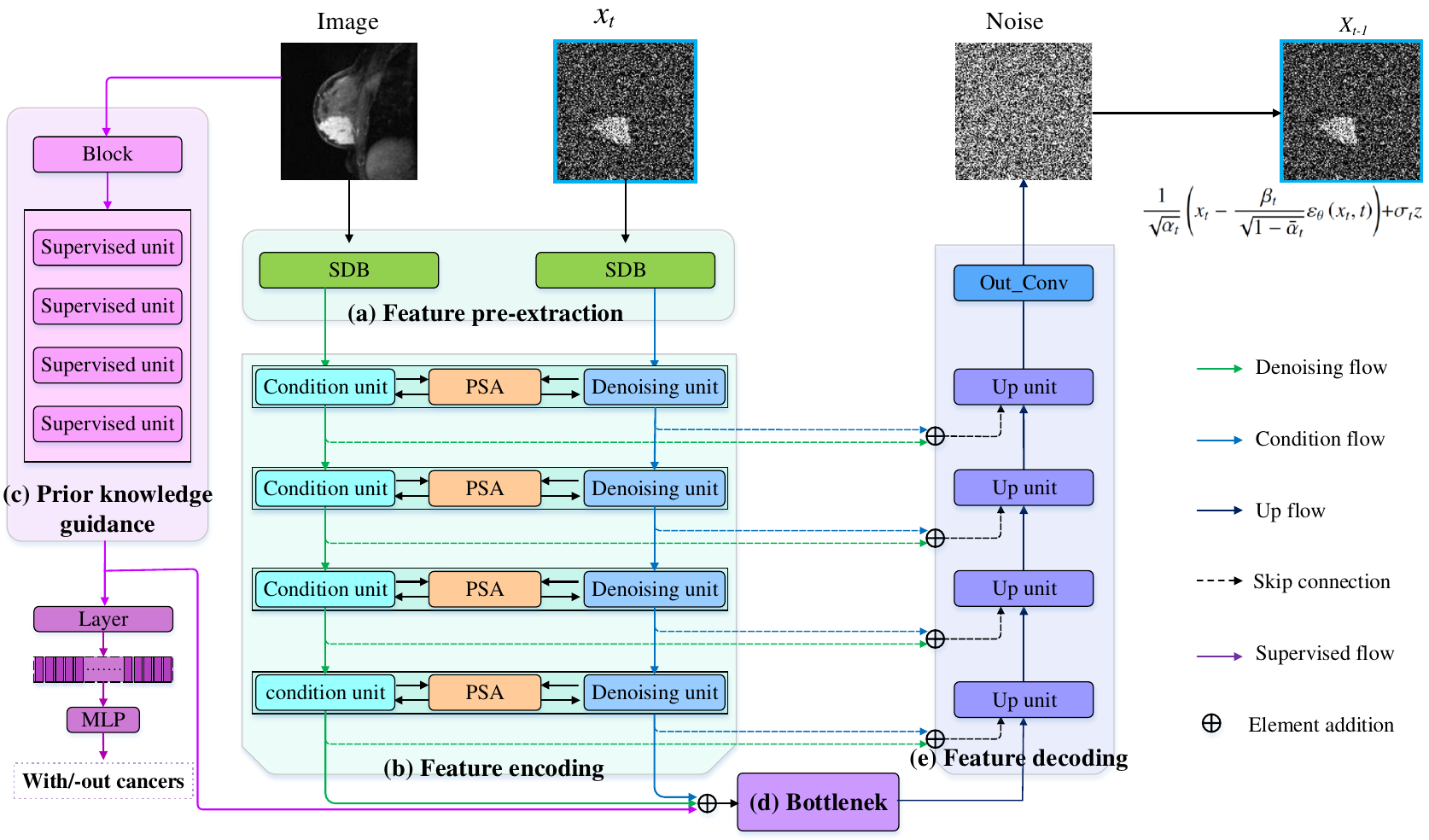}
\caption{Framework of PGDiffSeg model. 
\textbf{(a)} The \textbf{feature pre-extraction} module utilizes the slim dense block (SDB) to extract high-level semantic features from the image and $x_t$. \textbf{(b)} The \textbf{feature encoding} module performs denoising and encodes the features of $x_t$ and the image using condition units. The features are fused using parameter-shared attention (PSA) module after each layer of branches. \textbf{(c)} The \textbf{prior knowledge guidance} module plays a crucial role in the bottleneck layer by injecting expert knowledge. \textbf{(d)} The \textbf{bottleneck} layer serves as the connection point between feature encoding and feature decoding. It receives the output from feature encoding and passes it to feature decoding. \textbf{(e)} The \textbf{feature decoding} module converts the high-level feature representations from the bottleneck into the corresponding output for the original $x_t$. The model predicts the noise added in this step to obtain $x_{t-1}$.}
\label{fig2:Overall-view-of-our-model}
\end{figure*}

\subsection{Framework of the GFDiffSeg model}
Our proposed network GFDiffSeg operates on the diffusion process and generates segmented results from Gaussian noise through iterative sampling. The overall structure of the network is shown in Fig. \ref{fig2:Overall-view-of-our-model}. 

Our model comprises feature pre-extraction, feature encoding, prior knowledge guidance, bottleneck, and feature decoding stages. In the feature pre-extraction stage, the network separately extracts hidden features of the image and the noised label $x_t$ using two slim dense blocks (SDB). Next, in the feature encoding stage, the downflow of the model is divided into two branches. One branch, called the condition flow, takes the extracted image features as input to transmit spatial characteristics. And the other branch, the denoising flow, takes the extracted $x_t$ features as input to distinguish noise information. Unlike conventional encoders, these two extractor flows work in tandem, with the parameter-shared attention (PSA) module employed after each layer of the down-level unit, facilitating information exchange between the two branches and promoting mutual improvement. After feature encoding, the features from both sides are fused by element addition before the bottleneck. Furthermore, we also incorporate prior tumor information into the segmentation task by training a classification network that guides the learning process of the segmentation task. Following the bottleneck, the rich features are fed into a decoder consisting of four up units, ultimately obtaining the noise-adding prediction from $x_{t-1}$ to $x_t$.

Next, we will mainly introduce our three main tasks, SDB in feature pre-extraction, PAS module in feature encoding, and prior-supervision module used for prior knowledge guidance. We do not introduce too many innovations in other modules and have followed the majority's work; the detailed implementation can be found in the supplementary materials.

\subsection{Slim dense block}
\begin{figure}[htbp]
\centering
\includegraphics[scale=.52]{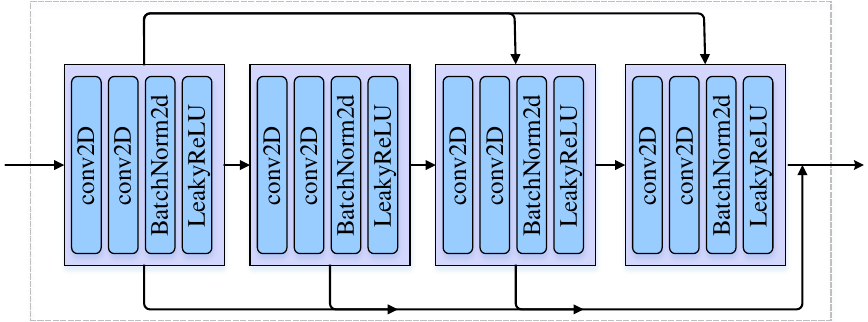}
\caption{The structure of the SDB as depicted in Fig. \ref{fig2:Overall-view-of-our-model}(a). Taking L=4 as an example, we showcase the flow of feature vectors between different modules. Different feature vectors are combined by addition, and for better visual effects, we do not reflect this process in the figure.}
\label{fig3:SDB}
\end{figure}

As shown in Fig \ref{fig3:SDB}, the slim dense block (SDB) is constructed using a lightweight dense block \cite{Huang_2017_CVPR}, and plays a crucial role in extracting hidden information that may be lost or inaccessible during the information exchange process of the PSA module in advance. By eliminating the growth rate of the Dense Block and adopting element-wise addition instead of channel concatenation, the SDB ensures a consistent number of channels in each layer. This design enables seamless integration of the SDB into our model, irrespective of the increasing number of layers.

This module consists of $L$ layers, where each layer implements a non-linear transformation $H(\cdot)$, which involves a composite function that includes the connection of two $3\times 3$ convolution blocks, two-dimensional BatchNorm, and LeakyReLU activation.

To fully exploit information, every layer in this module receives the output of all previous layers. This design ensures comprehensive information utilization and prevents overfitting by allowing the model to capture relevant features effectively. We denote the output of the $i^{th}$ layer as $x_i$. The transmission of feature maps in the $i^{th}$ layer can be represented as Equation \ref{eq101}.

\begin{equation}
x_{\mathrm{i}}=H\left(x_{i-1}\right) \oplus \sum_{j=1}^{i-1} x_{j} \circ \chi
\label{eq101}
\end{equation}
where $\chi$ is a scaling factor set to 0.2 in this model to optimize the overall architecture.

\subsection{Parameter-shared cross-attention module}
\label{sec3.4.2}
The parameter-shared cross-attention(PSA) module is designed to bridge the semantic gap between the two branches. Inspired by \cite{shaker2023unetr} and \cite{pmlr-v97-zhang19d}, we propose a novel approach for integrating relevant information, overcome the isolated state of the independent downflows between the two branches.

\begin{figure}[htbp]
\centering
\includegraphics[scale=.6]{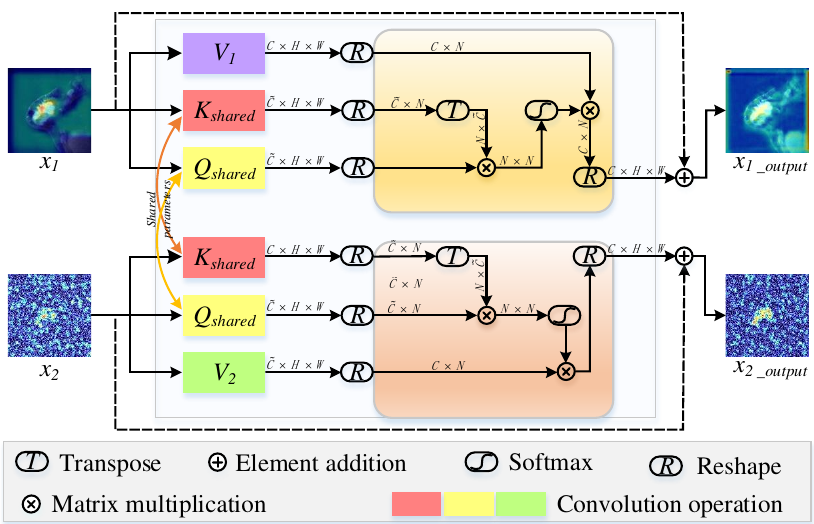}
\caption{Parameter-shared multi-layer cross-attention (PSA) module. After being embedded in the denoising unit and condition unit of each layer, this module receives the two intermediate results of the denoising flow and condition flow in the Feature encoder, fully fuses their related information, and then transmits them to the denoising unit and condition unit of the next layer respectively.}
\label{fig3:PSA}
\end{figure}

In our design, we introduce a PSA module after each layer of the denoising and condition units, as depicted in Fig. \ref{fig3:PSA}. Since $x_t$ and image have spatial consistency, the parameters of each layer's PSA module can be trained with two hidden features simultaneously: the feature that needs to be denoised and the pathological feature that contains tumor information. This PSA block plays a bidirectional role in embedding denoised features and pathological features into the same feature space, allowing them to complement each other.

The pathological features $x_1 \in R^{C \times H \times W}$ from the previous condition unit are transmitted to two semantic spaces for query and key calculations, which are mapped to the $R^{\tilde{C} \times N}$ space, where $\tilde{C}=\frac{C}{8}$ and $N=H\times W$. The denoised features $x_2$ also utilize the correspondent noise spaces with identical parameters, ensuring consistency in the computations and allowing one set of features to be mapped onto the other set. Then $x_1$ and $x_2$ use their separate value generation modules to obtain their value of $x_1$ and $x_2$ of dimensions $C\times N$ respectively. Therefore, the calculation of query, key, and value for the two features can be represented as Equation \ref{eq103}. 

\begin{equation}
q_i, k_i, v_i=Q_{shared }\left(x_i\right), K_{shared }\left(x_i\right), V_i\left(x_i\right);i=1,2
\label{eq103}
\end{equation}

where $Q$, $K$, $V$, are the generation of query, key, and value, respectively; and $q_i$, $k_i$, $v_i$ are query, key, and value of $x_i$.
The $x_1$ attention maps are first computed by multiplying the transpose of the projected q1 layer with k1. After measuring similarity through softmax, they are multiplied by $v_1$ to produce the final semantic maps with a shape of $N\times C$, which can be seen in Equation \ref{eq12}. Similarly, the denoising maps of $x_2$ undergo the same process as described in Equation \ref{eq13}.

\begin{equation}
X_s=v_1 \cdot {Softmax}\left(q_1^T \cdot k_1\right)
\label{eq12}
\end{equation}
\begin{equation}
X_d=v_2 \cdot \operatorname{Softmax}\left(q_2^T \cdot k_2\right)
\label{eq13}
\end{equation}
where $X_s$ and $X_d$ denote semantic and denoising maps, respectively.
Then the shape of $X_s$ and $X_d$ is reset to $C \times H \times W$, and the two original feature maps can be updated to form the output, as represented by Equation \ref{eq105}.

\begin{equation}
x_1=\eta_1 X_s+x_1; x_2=\eta_2 X_d+x_2
\label{eq105}
\end{equation}

where $\eta_1$ is a learnable scalar and we initialize it as 0. This allows the model to focus more on the intrinsic features of its branch in the early stages of training and gradually assign more weights to the supplemental features \cite{pmlr-v97-zhang19d}.

\subsection{Prior-supervision module}
\label{3.5}
In practical applications, achieving pixel-wise classification can be a challenging task, whereas determining the presence of a tumor is relatively simpler \cite{WANG2023102687}. Therefore, we incorporate a supervised flow generated by the prior supervision module as prior semantic knowledge to offer position guidance and integrate it into the denoising process.

\begin{figure}[htbp]
\centering
\includegraphics[scale=.5]{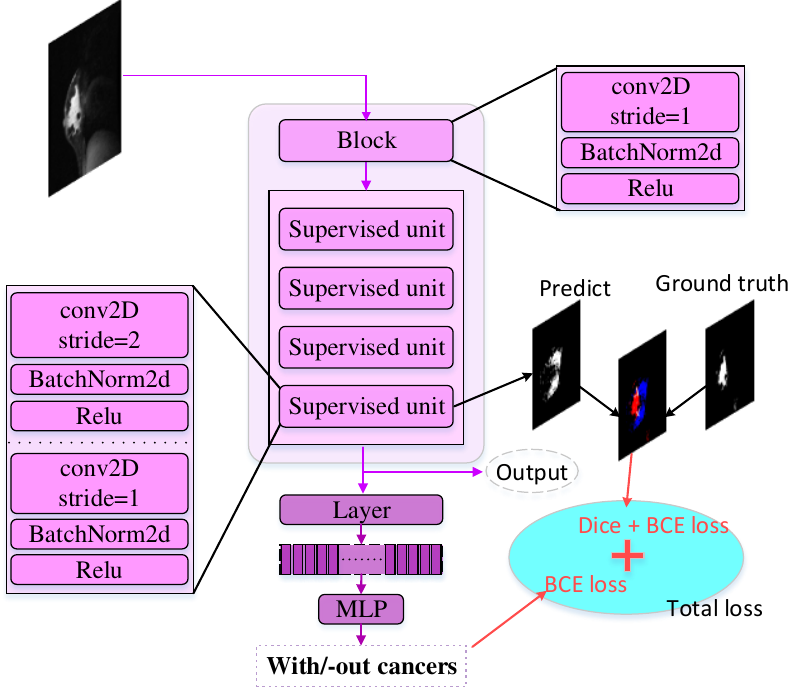}
\caption{Details of prior-supervision  module. To make the output of this module consistent with the Vector space of the output of the feature encoder, we designed one block and four supervised units, respectively, corresponding to the SDB and four layers units in the feature encoder 
}
\label{fig3:guidance}
\end{figure}

As shown in Fig. \ref{fig3:guidance}, this branch takes the image $I$ as input and first goes through a 3$\times$3 convolution layer, BatchNorm, and ReLU. Then, the output is fed through four supervised units, each consisting of two blocks with a 3$\times$3 convolution, BatchNorm, and ReLU. We use convolution with a stride of 2 to reduce the size of the feature maps, ensuring that the output size of the supervised flow matches the bottleneck of the main branch. Finally, the bottleneck-information weight is adjusted by adding x to $\tilde{x}$, depicted as Equation \ref{eq109}.

\begin{equation}
\tilde{x}=x \oplus \tilde{x}
\label{eq109}
\end{equation}
where $\tilde{x}$ represents the feature map of the bottleneck and $x$ is the output of the supervised flow. This module has an independent loss function. We obtain a probability classification value $p$ using linear mapping and train this part using cross-entropy loss, as shown in Equation \ref{eq110}.

\begin{equation}
{loss}_{CE}=-\frac{1}{N} \Sigma\left(y_i \log \left(p_i\right)+\left(1-y_i\right) \log \left(1-p_i\right)\right)
\label{eq110}
\end{equation}
where $N$ represents the number of samples, $y_i$ is the true value of the $i^\text{th}$ sample (whether a tumor exists), and $p_i$ is the predicted probability of that sample.

Furthermore, to ensure that this supervised flow focuses on the correct positions, we map the penultimate supervised unit from the space of $R^{C \times H \times W}$ to $R^{1 \times H \times W}$ and compute the Dice loss and Binary Cross-Entropy loss by comparing it with the resized segmentation label. The Dice loss and Binary Cross-Entropy loss are represented by Equation \ref{e22} and Equation \ref{e23}, respectively.

\begin{equation}
{loss}_{dice} = 1 - \frac{2 \sum\limits_{i=1}^{N} y_i \hat{y}_i + \epsilon}{\sum\limits_{i=1}^{N} y_i + \sum\limits_{i=1}^{N} \hat{y}_i + \epsilon}
\label{e22}
\end{equation}

\begin{equation}
{loss}_{BCE} = - \sum\limits_{i=1}^{N} y_i \log(\hat{y}_i) + (1 - y_i) \log(1 - \hat{y}_i)
\label{e23}
\end{equation}
where $y_i$ represents the true label, $\hat{y}_i$ represents the predicted result of the model, $N$ represents the number of samples, and $\epsilon$ is a small constant used to avoid division by zero. The final loss function is as Equation \ref{eq25}.

\begin{equation}
{loss} = \lambda_1 {loss}_{CE} + \lambda_2 ({{loss}_{dice} + {loss}_{BCE}})
\label{eq25}
\end{equation}


\section{Experiments}
\label{sec4}

\subsection{Datasets}
To demonstrate the effectiveness of tumor segmentation, we conduct comprehensive experiments on two publicly available datasets, the Breast-MRI-NACT-Pilot dataset \cite{doi:10.2214/ajr.184.6.01841774} and the Breast Ultrasound Image (BUSI) dataset \cite{ALDHABYANI2020104863}. 

\textbf{Breast-MRI-NACT-Pilot} is published on TCIA \cite{Clark2013} and used for locally advanced breast cancer segmentation. This collection contains 64 patients with stage 2 or 3 locally advanced breast cancer (tumor size $>=$ 3cm) enrolled from 1995 to 2002. All patients had breast cancer diagnosed based on histopathological reports of biopsy or surgical excision. Using a bilateral phased array breast coil, the breast MRI was constructed by a 1.5-T scanner (Signa, GE Healthcare, Milwaukee, WI). Our experiment obtained DCE-MRI scans of 64 patients before NACT treatment, of which 62 scans have a resolution of 256 × 256 × 60, and the resolution of the two scans is 512x512x64. We trained our model using 2D slices, with 43 patients for training, 7 for validation, and 14 for testing.

\textbf{BUSI} is the first publicly available breast ultrasound dataset. It was collected in 2018 using the LOGIQ E9 ultrasound system and LOGIQ E90 Agile ultrasound system during the scanning process, including breast ultrasound images from 600 women between the ages of 25 and 75. The dataset comprises 780 grayscale images with an average size of 500 $\times$ 500 pixels. These images were collected and stored in DICOM format at Baheya Hospital and then converted to PNG format using a DICOM converter application. The collected images are divided into three categories: normal (133 images), benign(437 images), and malignant(210 images).

\subsection{Data preprocessing}
In the Breast-MRI-NACT-Pilot dataset, approximately the first 70 \% of patients in each dataset are used for training, the middle 10\% for validation, and the remaining 20\% for testing. We applied the same partitioning for the BUSI dataset's three classes (normal, benign, and malignant) to ensure consistency in sample distribution across the training, validation, and testing sets. We truncate all pixel intensities to a specific range. In the Breast-MRI-NACT-Pilot dataset, the original HU range is from 0 to 96 but can vary for each patient. We normalize each patient's range to [0,255] and then truncate the intensity values to the range of [20, 200]. For the BUSI dataset, the original distribution is [0, 255], and we truncate it to the range of [30, 235]. To prevent overfitting, data augmentation, including horizontal, vertical flipping, and rotation by 90°, 180°, and 270°, were applied to the BUSI dataset to increase the size of the training set six-fold. We resized all data to $128 \times 128$ and normalized it to [-1,1] based on the methodology proposed in DDPM \cite{NEURIPS2020_4c5bcfec}.

\begin{figure*}[!t]
\centering
\includegraphics[scale=.5]{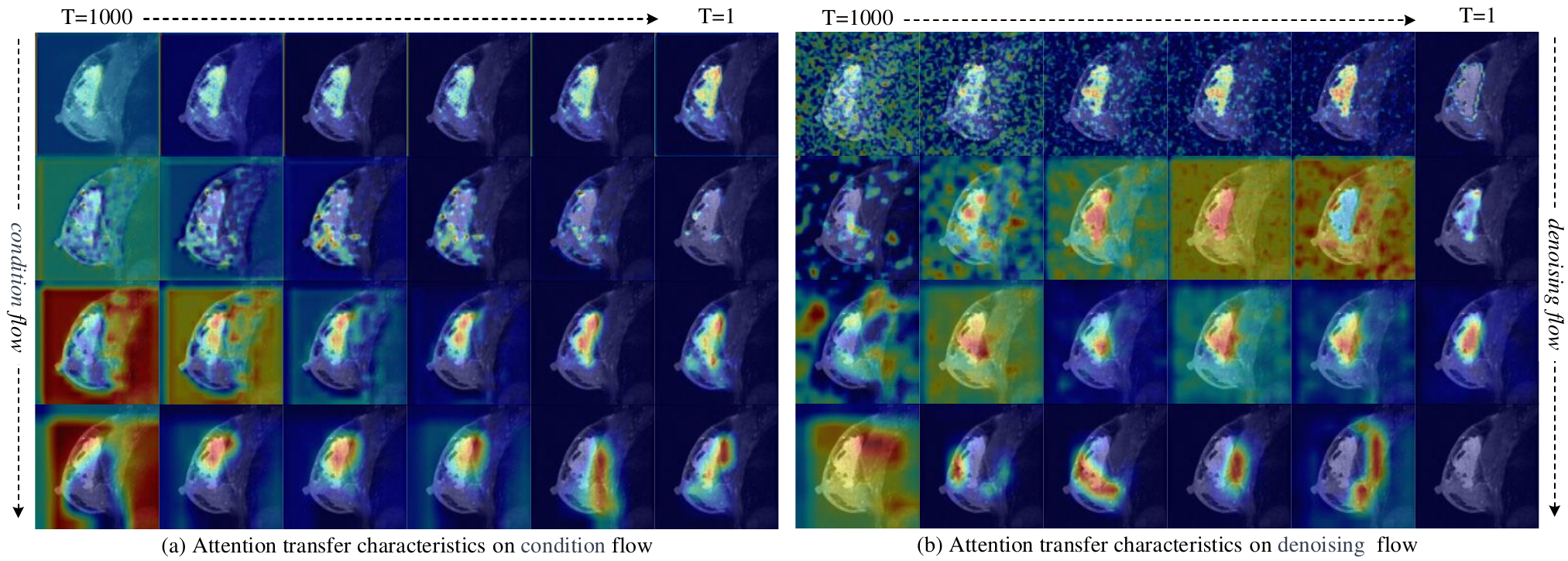}
\caption{The attention positions in the (a) condition flow and (b) denoising flow exhibit variations across diffusion steps. The four rows correspond to the four layers of the model. Within each row, the images represent the model's output at the same position but at different diffusion steps. Horizontal arrows illustrate the changes in diffusion steps, while vertical arrows indicate the direction of parameter propagation within the model. The original images and their corresponding ground truth are displayed on Fig. \ref{usage-of-attention}(c). The highlighted red regions signify the positions significantly influencing the model's decision-making process.}
\label{down}
\end{figure*}

\section{Discussion}
\label{sec6}
In this work, we demonstrate the effectiveness and clinical feasibility of the proposed method in improving cancer segmentation, which includes a visual analysis and discussion of the proposed modules, as well as the applicability of existing efficient sampling studies to our tumor segmentation task.
To directly reflect the execution process of our model, we employed Grad-CAM \cite{Selvaraju_2017_ICCV} to highlight the regions in each layer that contribute to the segmentation results.

\subsection{Attention transfer characteristics of denoising model}
\label{sec6-1}
We demonstrate the attention transfer characteristics of the denoising model in the segmentation task, as shown in Fig. \ref{down}. It illustrates the spatiotemporal transition of the vital contribution regions when generating a segmentation mask. In Fig. \ref{down}(a), for example, each row represents one layer's output of the model's condition flow, with diffusion steps decreasing from left to right and the model's depth increasing from top to bottom.

In the same layer in condition flow in Fig. \ref{down}(a), attention positions vary at different diffusion steps. This variation often transitions smoothly as the timesteps decrease, but occasional jumps may occur. Additionally, we observe that the first layer of the condition flow initially focuses on global information and gradually localizes to the tumor region. The attention in the second layer appears more dispersed, similar to the distribution shown in Fig. \ref{down}(b), which may be attributed to the influence of the sudden introduction of information from the denoising flow. However, as the model's depth increases, this bias is readjusted. The attention in the third layer shifts gradually from the background to the foreground. In the deeper layer (the last row), the model's attention becomes diverse, indicating that the deep network attempts to learn more complex features that cannot be easily interpreted.

Similarly, Fig. \ref{down}(b) depicts the changes in attention positions of the denoising flow. We present the results for each layer, allowing us to observe the variation in attention positions over time. Fig. \ref{down}(b) demonstrates that the denoising flow exhibits a similar pattern to the condition flow, indicating that our model operates in a rich feature space. Although this branch's initial input is Gaussian noise (at $T$=1000), it can still accurately locate the tumor when $T$ gradually decreases. This ability can be attributed to introducing PSA.

\subsection{Effectiveness of PSA Module}
\label{sec6-2}
\begin{figure}[htbp]
\centering
\includegraphics[scale=.45]{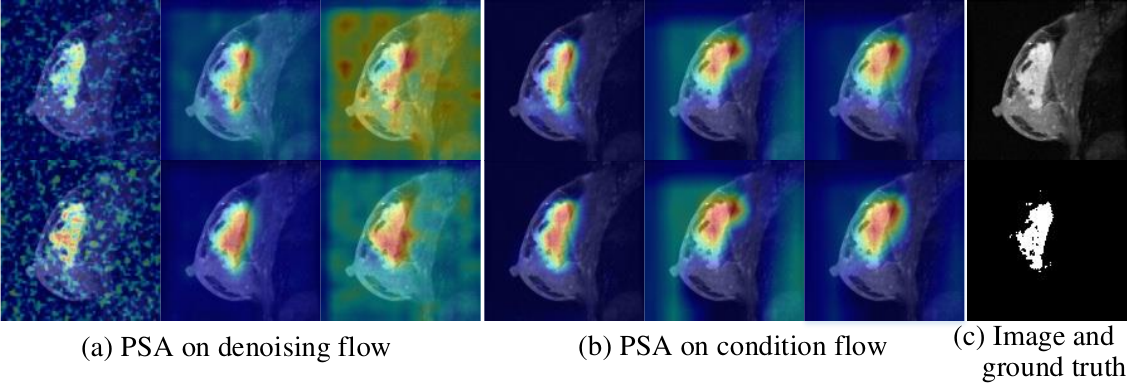}
\caption{The positions attended by the convolutional layers before and after PSA are shown in the top and bottom rows, respectively. (a) represents the results from the denoising flow, while (b) represents the results from the condition flow. Moreover, (c) is the original images and their corresponding ground truth.}
\label{usage-of-attention}
\end{figure}

As shown in Fig. \ref{usage-of-attention}, we compared the changes in the heatmaps before and after applying the PSA module, where the top row represents the attention positions of the latest convolutional layer before PSA, and the bottom row displays the attention maps after PSA. In Fig. \ref{usage-of-attention}(a), we notice that the model focuses more on the tumor region after incorporating the PSA module in the denoising flow. This change is crucial because we need PSA to introduce semantic information from the condition flow, enabling it to perform denoising effectively in this step. Additionally, Fig. \ref{usage-of-attention}(b) demonstrates that although the condition flow already captures the semantic information and focuses on the tumor region, introducing the PSA module further enhances its confidence in the target area.

\subsection{The effectiveness of introducing prior knowledge}
\label{sec6-3}
\begin{figure}[htbp]
\centering
\includegraphics[scale=.45]{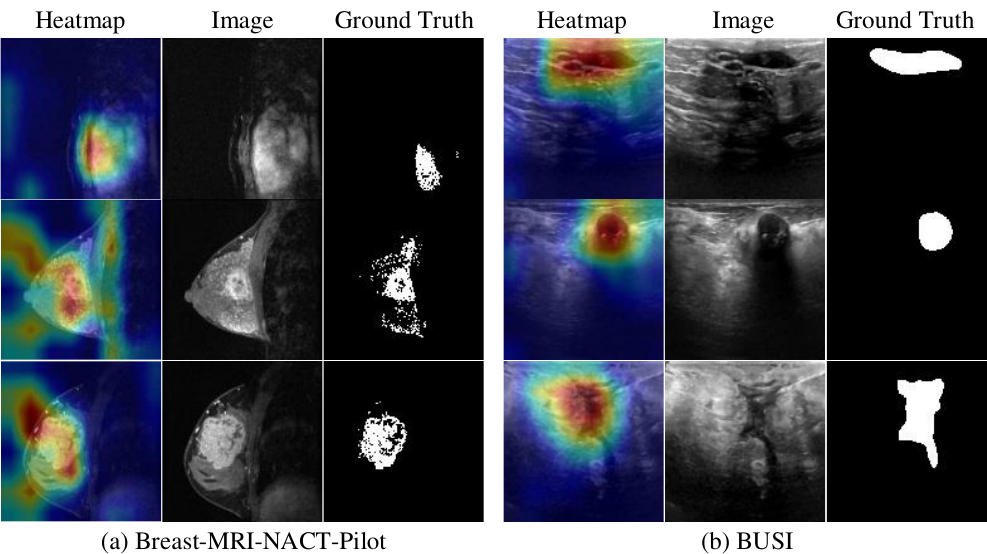}
\caption{The effectiveness of introducing prior knowledge. The heatmaps indicate the model's greater attention at the highlighted positions, represented by red.}
\label{prior_knowledge_cam}
\end{figure}

In this section, we discussed the effectiveness of prior knowledge. We examined the attention positions of the last layer feature map in this branch using CAM, as shown in Fig. \ref{prior_knowledge_cam}. It can be observed that this subnetwork can identify the location of tumors, which means it can simulate prior knowledge provided by physicians and facilitate subsequent denoising. In the Breast-MRI-NACT-Pilot dataset, tumor locations are often scattered. Although this subnetwork cannot offer precise segmentation like the segmentation network, it can at least identify the approximate location of the tumor, especially in areas where tumors are concentrated. While it's not absolutely accurate, introducing this prior knowledge provides convenience for the leading network.

\subsection{Analyzing of the results}
\label{sec6-4}
\begin{figure}[!t]
\centering
\includegraphics[scale=.45]{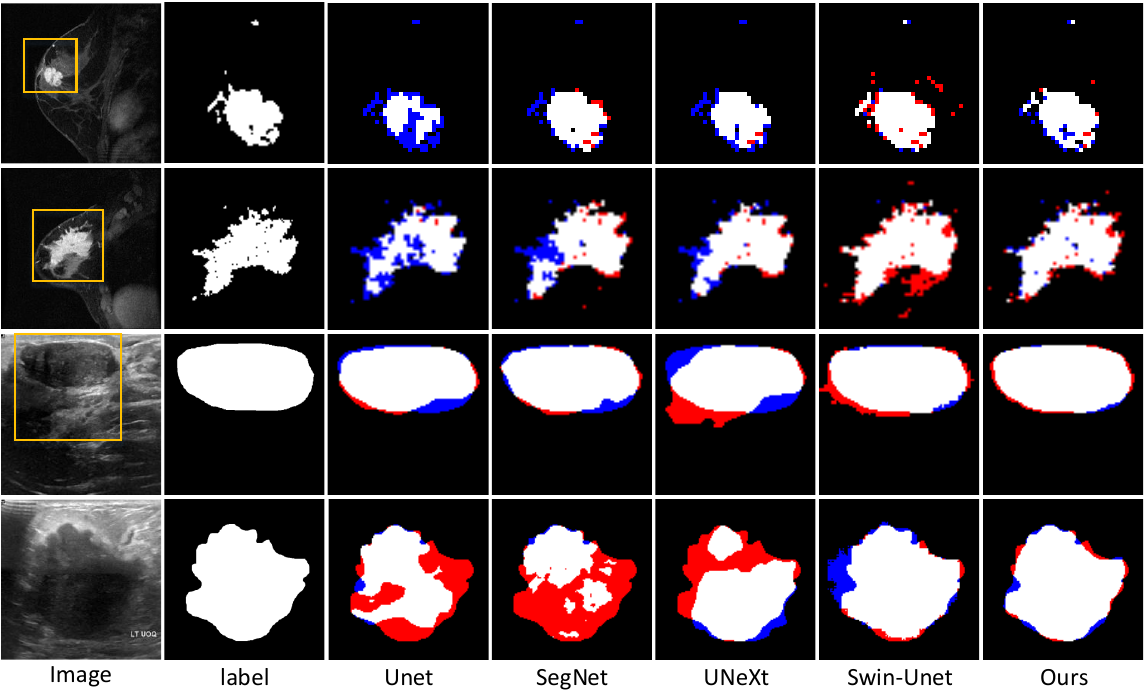}
\caption{Comparison of the GFDiffSeg model with state-of-the-art models. We use different colors to represent the results of the models in terms of correct segmentation, over-segmentation, and unsegmented regions. Using A and B to denote the ground truth region and the predicted label of the tumor, white represents A$\cap$B (intersection of A and B); red represents A-B, indicating the areas that the model failed to segment; blue represents B-A, showing regions that do not contain tumors but were incorrectly segmented as tumors by the model.}
\label{SOTA_comparation}
\end{figure}

Fig \ref{SOTA_comparation} compares the performance of the GFDiffSeg model and others. The first two rows show the results of different models on the Breast-MRI-NACT-Pilot dataset, while the last two rows depict the results on the BUSI dataset. It can be observed that other models often exhibit more regions of over-segmentation or unsegmented areas, whereas our model's inaccurate position is primarily located at the edges. Furthermore, both datasets indicate that Swin-Unet is prone to excessive segmentation and exhibits jagged edges.
Moreover, SegNet and UNeXt struggle with segmenting small detached tumors. The segmentation results of Swin-Unet tend to include some unrelated regions. On the BUSI dataset, as depicted in the last row of Fig. \ref{SOTA_comparation}, Unet and SegNet cannot fully segment tumors with a significant difference in the grayscale range. UNeXt, on the other hand, fails to identify the intersection between two grayscale ranges as tumor regions. Fortunately, this problem is effectively resolved in GFDiffSeg, as the denoising diffusion model is powerful for learning data distributions.

\subsection{Efficient sampling}
\begin{figure}[htbp]
\centering
\includegraphics[scale=.17]{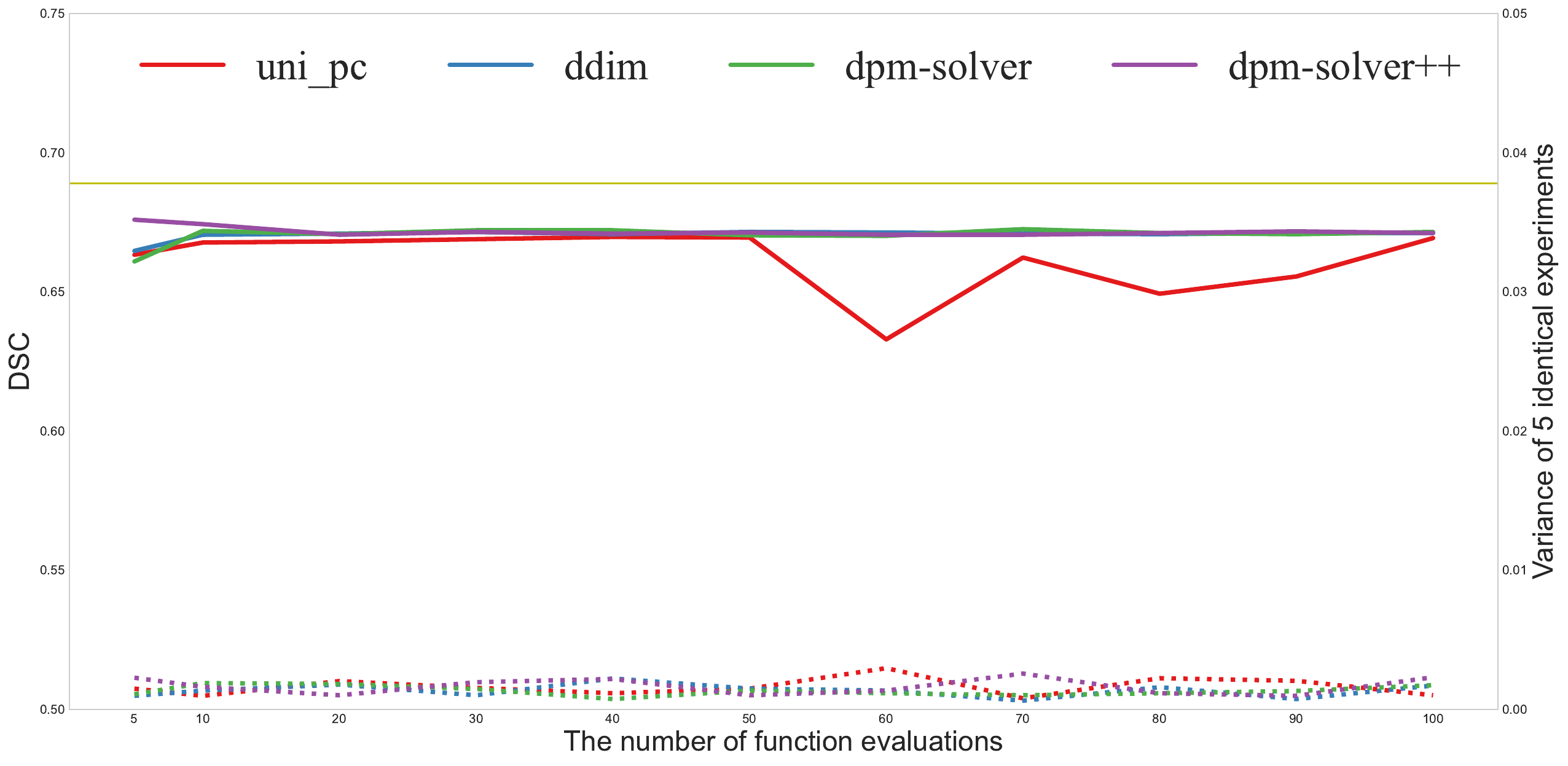}
\caption{The mean and variance of the DSC for different numbers of function evaluations on the Breast-MRI-NACT-Pilot dataset.}
\label{fig:NACT_solver}
\end{figure}

\begin{figure}[htbp]
\centering
\includegraphics[scale=.17]{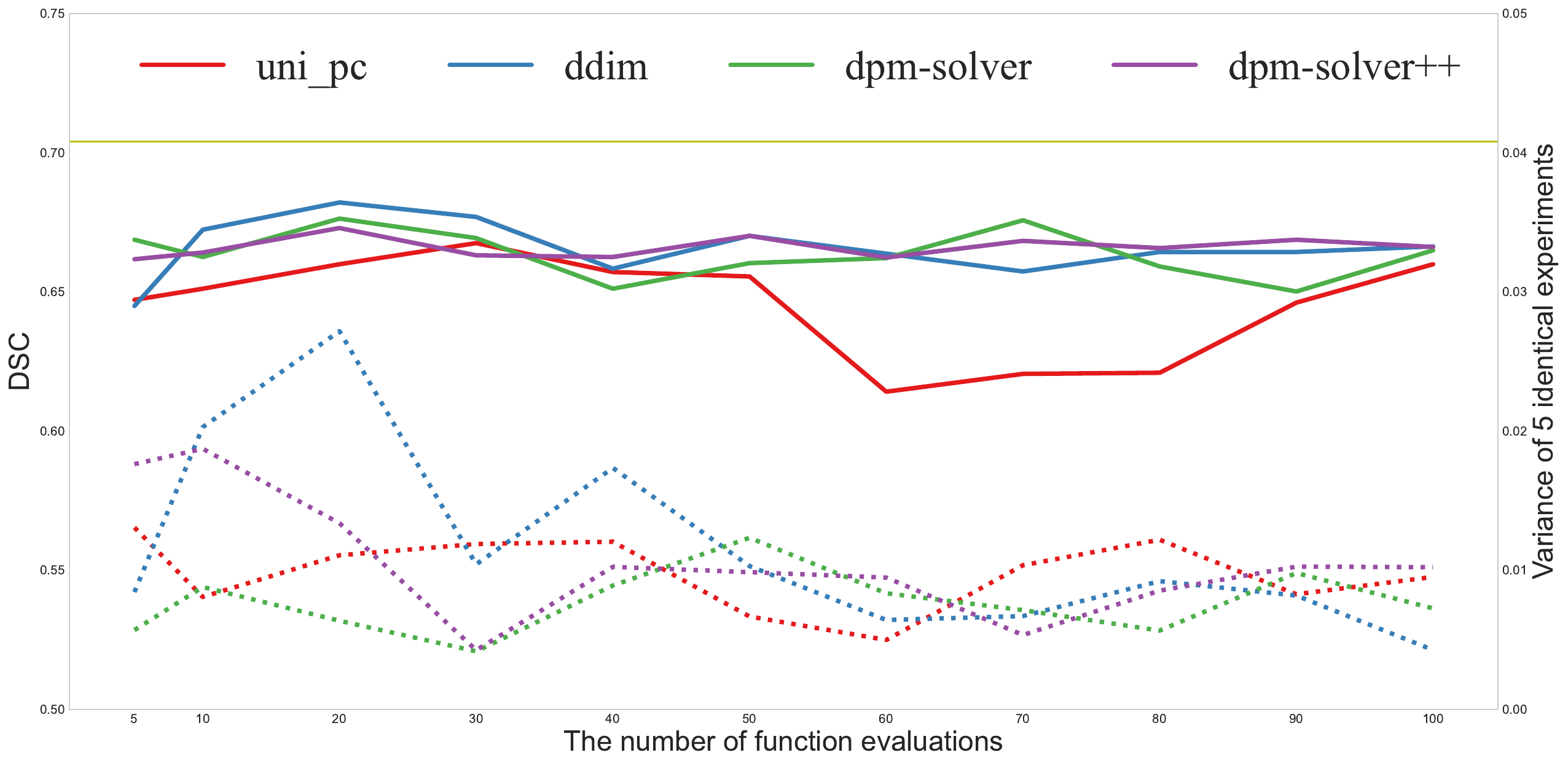}
\caption{The mean and variance of the DSC for different numbers of function evaluations on the BUSI dataset.}
\label{fig:BUSI_solver}
\end{figure}

Since diffusion models require extensive sampling iterations, which significantly hinders their application, we employed accelerated algorithms \cite{song2021denoising, NEURIPS2022_260a14ac, lu2022dpm, NEURIPS2023_9c2aa1e4} to explore efficient sampling with reduced numbers of function evaluations (NFE). To mitigate the impact of randomness, we repeated each experiment five times and reported the average and variance of the Dice similarity score (DSC). More details can be found in the supplemental materials. The visualized results are shown in Figures \ref{fig:NACT_solver} and \ref{fig:BUSI_solver}, where the solid line represents the average DSC of the five repeated experiments, the dashed line represents the variance, and the yellow horizontal line indicates the initial results of our model.


\section{Conclusion}
\label{sec7}
We propose the GFDiffSeg model, a novel approach for breast cancer segmentation. Our study further demonstrates the feasibility of the denoising strategy in medical image segmentation, specifically tailored for breast cancer. We enhance tumor localization by incorporating the PSA module and a guided prior knowledge strategy. Experimental results show that the GFDiffSeg model outperforms or performs comparably to state-of-the-art approaches. Interpretability analysis further confirms the effectiveness of our method.

However, there are limitations in the current version of the GFDiffSeg model that need to be addressed in future research. We have only conducted experiments on breast datasets with different modalities. The denoising process also incurs longer sampling times than other models, limiting the practical application and generalization of the denoising diffusion model in medical image segmentation.

In the future, We aim to explore methods to reduce sampling time and extend our model to tumor segmentation in other organs. Additionally, we will focus on developing more efficient model architectures, such as leveraging prior knowledge to design prompts that guide the model in generating more accurate segmentation results.

\section*{Acknowledgment}
This work is supported by National Nature Science Foundation of China (62072290), Natural Science Foundation of Shandong Province (ZR2022QF022) and Jinan "20 new colleges and universities" Funded Project (202228110). 

\bibliographystyle{IEEEtran}
\bibliography{template}

\begin{thebibliography}{10}
\providecommand{\url}[1]{#1}
\csname url@samestyle\endcsname
\providecommand{\newblock}{\relax}
\providecommand{\bibinfo}[2]{#2}
\providecommand{\BIBentrySTDinterwordspacing}{\spaceskip=0pt\relax}
\providecommand{\BIBentryALTinterwordstretchfactor}{4}
\providecommand{\BIBentryALTinterwordspacing}{\spaceskip=\fontdimen2\font plus
\BIBentryALTinterwordstretchfactor\fontdimen3\font minus \fontdimen4\font\relax}
\providecommand{\BIBforeignlanguage}[2]{{%
\expandafter\ifx\csname l@#1\endcsname\relax
\typeout{** WARNING: IEEEtran.bst: No hyphenation pattern has been}%
\typeout{** loaded for the language `#1'. Using the pattern for}%
\typeout{** the default language instead.}%
\else
\language=\csname l@#1\endcsname
\fi
#2}}
\providecommand{\BIBdecl}{\relax}
\BIBdecl

\bibitem{https://doi.org/10.3322/caac.21763}
\BIBentryALTinterwordspacing
R.~L. Siegel, K.~D. Miller, N.~S. Wagle, and A.~Jemal, ``Cancer statistics, 2023,'' \emph{CA: A Cancer Journal for Clinicians}, vol.~73, no.~1, pp. 17--48, 2023. [Online]. Available: \url{https://acsjournals.onlinelibrary.wiley.com/doi/abs/10.3322/caac.21763}
\BIBentrySTDinterwordspacing

\bibitem{doi:10.1126/science.aay9040}
\BIBentryALTinterwordspacing
D.~Crosby, S.~Bhatia, K.~M. Brindle, L.~M. Coussens, C.~Dive, M.~Emberton, S.~Esener, R.~C. Fitzgerald, S.~S. Gambhir, P.~Kuhn, T.~R. Rebbeck, and S.~Balasubramanian, ``Early detection of cancer,'' \emph{Science}, vol. 375, no. 6586, p. eaay9040, 2022. [Online]. Available: \url{https://www.science.org/doi/abs/10.1126/science.aay9040}
\BIBentrySTDinterwordspacing

\bibitem{CROSWELL2010202}
\BIBentryALTinterwordspacing
J.~M. Croswell, D.~F. Ransohoff, and B.~S. Kramer, ``Principles of cancer screening: Lessons from history and study design issues,'' \emph{Seminars in Oncology}, vol.~37, no.~3, pp. 202--215, 2010, cancer Prevention I. [Online]. Available: \url{https://www.sciencedirect.com/science/article/pii/S0093775410000710}
\BIBentrySTDinterwordspacing

\bibitem{https://doi.org/10.1002/ijc.29634}
\BIBentryALTinterwordspacing
T.~J. O'Grady, M.~A. Gates, and F.~P. Boscoe, ``Thyroid cancer incidence attributable to overdiagnosis in the united states 1981–2011,'' \emph{International Journal of Cancer}, vol. 137, no.~11, pp. 2664--2673, 2015. [Online]. Available: \url{https://onlinelibrary.wiley.com/doi/abs/10.1002/ijc.29634}
\BIBentrySTDinterwordspacing

\bibitem{SASAKI2015e186}
\BIBentryALTinterwordspacing
K.~Sasaki, S.~S. Strom, S.~O'Brien, E.~Jabbour, F.~Ravandi, M.~Konopleva, G.~Borthakur, N.~Pemmaraju, N.~Daver, P.~Jain, S.~Pierce, H.~Kantarjian, and J.~E. Cortes, ``Relative survival in patients with chronic-phase chronic myeloid leukaemia in the tyrosine-kinase inhibitor era: analysis of patient data from six prospective clinical trials,'' \emph{The Lancet Haematology}, vol.~2, no.~5, pp. e186--e193, 2015. [Online]. Available: \url{https://www.sciencedirect.com/science/article/pii/S2352302615000484}
\BIBentrySTDinterwordspacing

\bibitem{5656974}
L.~Singh, Z.~Jaffery, Z.~Zaheeruddin, and R.~Singh, ``Segmentation and characterization of breast tumor in mammograms,'' in \emph{2010 International Conference on Advances in Recent Technologies in Communication and Computing}, 2010, pp. 213--216.

\bibitem{10130343}
J.~Yang, L.~Jiao, R.~Shang, X.~Liu, R.~Li, and L.~Xu, ``Ept-net: Edge perception transformer for 3d medical image segmentation,'' \emph{IEEE Transactions on Medical Imaging}, pp. 1--1, 2023.

\bibitem{9154595}
Z.~Tian, H.~Zhao, M.~Shu, Z.~Yang, R.~Li, and J.~Jia, ``Prior guided feature enrichment network for few-shot segmentation,'' \emph{IEEE Transactions on Pattern Analysis and Machine Intelligence}, vol.~44, no.~2, pp. 1050--1065, 2022.

\bibitem{9356353}
S.~Minaee, Y.~Boykov, F.~Porikli, A.~Plaza, N.~Kehtarnavaz, and D.~Terzopoulos, ``Image segmentation using deep learning: A survey,'' \emph{IEEE Transactions on Pattern Analysis and Machine Intelligence}, vol.~44, no.~7, pp. 3523--3542, 2022.

\bibitem{9669083}
J.-J. Liu, Q.~Hou, Z.-A. Liu, and M.-M. Cheng, ``Poolnet+: Exploring the potential of pooling for salient object detection,'' \emph{IEEE Transactions on Pattern Analysis and Machine Intelligence}, vol.~45, no.~1, pp. 887--904, 2023.

\bibitem{8379359}
X.~Li, H.~Chen, X.~Qi, Q.~Dou, C.-W. Fu, and P.-A. Heng, ``H-denseunet: Hybrid densely connected unet for liver and tumor segmentation from ct volumes,'' \emph{IEEE Transactions on Medical Imaging}, vol.~37, no.~12, pp. 2663--2674, 2018.

\bibitem{10.1007/978-3-030-66415-2_27}
R.~Ke, A.~Bugeau, N.~Papadakis, P.~Schuetz, and C.-B. Sch{\"o}nlieb, ``Learning to segment microscopy images with lazy labels,'' in \emph{Computer Vision -- ECCV 2020 Workshops}, A.~Bartoli and A.~Fusiello, Eds.\hskip 1em plus 0.5em minus 0.4em\relax Cham: Springer International Publishing, 2020, pp. 411--428.

\bibitem{9444895}
Y.~Liu, F.~Zhang, C.~Chen, S.~Wang, Y.~Wang, and Y.~Yu, ``Act like a radiologist: Towards reliable multi-view correspondence reasoning for mammogram mass detection,'' \emph{IEEE Transactions on Pattern Analysis and Machine Intelligence}, vol.~44, no.~10, pp. 5947--5961, 2022.

\bibitem{LIU2023105145}
\BIBentryALTinterwordspacing
T.~Liu, H.~Wang, S.~Yu, F.~Feng, and J.~Zhao, ``A soft-attention guidance stacked neural network for neoadjuvant chemotherapy’s pathological response diagnosis using breast dynamic contrast-enhanced mri,'' \emph{Biomedical Signal Processing and Control}, vol.~86, p. 105145, 2023. [Online]. Available: \url{https://www.sciencedirect.com/science/article/pii/S1746809423005785}
\BIBentrySTDinterwordspacing

\bibitem{NEURIPS2020_4c5bcfec}
\BIBentryALTinterwordspacing
J.~Ho, A.~Jain, and P.~Abbeel, ``Denoising diffusion probabilistic models,'' in \emph{Advances in Neural Information Processing Systems}, H.~Larochelle, M.~Ranzato, R.~Hadsell, M.~Balcan, and H.~Lin, Eds., vol.~33.\hskip 1em plus 0.5em minus 0.4em\relax Curran Associates, Inc., 2020, pp. 6840--6851. [Online]. Available: \url{https://proceedings.neurips.cc/paper_files/paper/2020/file/4c5bcfec8584af0d967f1ab10179ca4b-Paper.pdf}
\BIBentrySTDinterwordspacing

\bibitem{9785606}
P.~Li, Y.~Liu, Z.~Cui, F.~Yang, Y.~Zhao, C.~Lian, and C.~Gao, ``Semantic graph attention with explicit anatomical association modeling for tooth segmentation from cbct images,'' \emph{IEEE Transactions on Medical Imaging}, vol.~41, no.~11, pp. 3116--3127, 2022.

\bibitem{9455423}
H.~Huang, H.~Zheng, L.~Lin, M.~Cai, H.~Hu, Q.~Zhang, Q.~Chen, Y.~Iwamoto, X.~Han, Y.-W. Chen, and R.~Tong, ``Medical image segmentation with deep atlas prior,'' \emph{IEEE Transactions on Medical Imaging}, vol.~40, no.~12, pp. 3519--3530, 2021.

\bibitem{gong2022pet}
K.~Gong, K.~Johnson, G.~El~Fakhri, Q.~Li, and T.~Pan, ``Pet image denoising based on denoising diffusion probabilistic model,'' \emph{European Journal of Nuclear Medicine and Molecular Imaging}, vol.~51, no.~2, pp. 358--368, 2024.

\bibitem{CHUNG2022102479}
\BIBentryALTinterwordspacing
H.~Chung and J.~C. Ye, ``Score-based diffusion models for accelerated mri,'' \emph{Medical Image Analysis}, vol.~80, p. 102479, 2022. [Online]. Available: \url{https://www.sciencedirect.com/science/article/pii/S1361841522001268}
\BIBentrySTDinterwordspacing

\bibitem{amit2022segdiff}
\BIBentryALTinterwordspacing
T.~Amit, T.~Shaharbany, E.~Nachmani, and L.~Wolf, ``Segdiff: Image segmentation with diffusion probabilistic models,'' 2021. [Online]. Available: \url{https://arxiv.org/abs/2112.00390}
\BIBentrySTDinterwordspacing

\bibitem{zimmermann2021scorebased}
R.~S. Zimmermann, L.~Schott, Y.~Song, B.~A. Dunn, and D.~A. Klindt, ``Score-based generative classifiers,'' in \emph{NeurIPS 2021 Workshop on Deep Generative Models and Downstream Applications}, 2021.

\bibitem{10.1007/978-3-031-16431-6_51}
B.~Kim and J.~C. Ye, ``Diffusion deformable model for 4d temporal medical image generation,'' in \emph{Medical Image Computing and Computer Assisted Intervention -- MICCAI 2022}, L.~Wang, Q.~Dou, P.~T. Fletcher, S.~Speidel, and S.~Li, Eds.\hskip 1em plus 0.5em minus 0.4em\relax Cham: Springer Nature Switzerland, 2022, pp. 539--548.

\bibitem{Moghadam_2023_WACV}
P.~A. Moghadam, S.~Van~Dalen, K.~C. Martin, J.~Lennerz, S.~Yip, H.~Farahani, and A.~Bashashati, ``A morphology focused diffusion probabilistic model for synthesis of histopathology images,'' in \emph{Proceedings of the IEEE/CVF Winter Conference on Applications of Computer Vision (WACV)}, January 2023, pp. 2000--2009.

\bibitem{10081412}
F.-A. Croitoru, V.~Hondru, R.~T. Ionescu, and M.~Shah, ``Diffusion models in vision: A survey,'' \emph{IEEE Transactions on Pattern Analysis and Machine Intelligence}, pp. 1--20, 2023.

\bibitem{pmlr-v172-wolleb22a}
\BIBentryALTinterwordspacing
J.~Wolleb, R.~Sandk\"uhler, F.~Bieder, P.~Valmaggia, and P.~C. Cattin, ``Diffusion models for implicit image segmentation ensembles,'' in \emph{Proceedings of The 5th International Conference on Medical Imaging with Deep Learning}, ser. Proceedings of Machine Learning Research, E.~Konukoglu, B.~Menze, A.~Venkataraman, C.~Baumgartner, Q.~Dou, and S.~Albarqouni, Eds., vol. 172.\hskip 1em plus 0.5em minus 0.4em\relax PMLR, 06--08 Jul 2022, pp. 1336--1348. [Online]. Available: \url{https://proceedings.mlr.press/v172/wolleb22a.html}
\BIBentrySTDinterwordspacing

\bibitem{wu2024medsegdiff}
J.~Wu, R.~Fu, H.~Fang, Y.~Zhang, Y.~Yang, H.~Xiong, H.~Liu, and Y.~Xu, ``Medsegdiff: Medical image segmentation with diffusion probabilistic model,'' in \emph{Medical Imaging with Deep Learning}.\hskip 1em plus 0.5em minus 0.4em\relax PMLR, 2024, pp. 1623--1639.

\bibitem{chowdary2023diffusion}
G.~J. Chowdary and Z.~Yin, ``Diffusion transformer u-net for medical image segmentation,'' in \emph{International conference on medical image computing and computer-assisted intervention}.\hskip 1em plus 0.5em minus 0.4em\relax Springer, 2023, pp. 622--631.

\bibitem{10.1007/978-3-030-87193-2_8}
D.~Karimi, S.~D. Vasylechko, and A.~Gholipour, ``Convolution-free medical image segmentation using transformers,'' in \emph{Medical Image Computing and Computer Assisted Intervention -- MICCAI 2021}, M.~de~Bruijne, P.~C. Cattin, S.~Cotin, N.~Padoy, S.~Speidel, Y.~Zheng, and C.~Essert, Eds.\hskip 1em plus 0.5em minus 0.4em\relax Cham: Springer International Publishing, 2021, pp. 78--88.

\bibitem{10.1007/978-3-031-25066-8_9}
H.~Cao, Y.~Wang, J.~Chen, D.~Jiang, X.~Zhang, Q.~Tian, and M.~Wang, ``Swin-unet: Unet-like pure transformer for medical image segmentation,'' in \emph{Computer Vision -- ECCV 2022 Workshops}, L.~Karlinsky, T.~Michaeli, and K.~Nishino, Eds.\hskip 1em plus 0.5em minus 0.4em\relax Cham: Springer Nature Switzerland, 2023, pp. 205--218.

\bibitem{10.1007/978-3-030-87193-2_4}
J.~M.~J. Valanarasu, P.~Oza, I.~Hacihaliloglu, and V.~M. Patel, ``Medical transformer: Gated axial-attention for medical image segmentation,'' in \emph{Medical Image Computing and Computer Assisted Intervention -- MICCAI 2021}, M.~de~Bruijne, P.~C. Cattin, S.~Cotin, N.~Padoy, S.~Speidel, Y.~Zheng, and C.~Essert, Eds.\hskip 1em plus 0.5em minus 0.4em\relax Cham: Springer International Publishing, 2021, pp. 36--46.

\bibitem{10.1007/978-3-030-87193-2_2}
Y.~Zhang, H.~Liu, and Q.~Hu, ``Transfuse: Fusing transformers and cnns for medical image segmentation,'' in \emph{Medical Image Computing and Computer Assisted Intervention -- MICCAI 2021}, M.~de~Bruijne, P.~C. Cattin, S.~Cotin, N.~Padoy, S.~Speidel, Y.~Zheng, and C.~Essert, Eds.\hskip 1em plus 0.5em minus 0.4em\relax Cham: Springer International Publishing, 2021, pp. 14--24.

\bibitem{9785614}
A.~Lin, B.~Chen, J.~Xu, Z.~Zhang, G.~Lu, and D.~Zhang, ``Ds-transunet: Dual swin transformer u-net for medical image segmentation,'' \emph{IEEE Transactions on Instrumentation and Measurement}, vol.~71, pp. 1--15, 2022.

\bibitem{Liu_2021_ICCV}
Z.~Liu, Y.~Lin, Y.~Cao, H.~Hu, Y.~Wei, Z.~Zhang, S.~Lin, and B.~Guo, ``Swin transformer: Hierarchical vision transformer using shifted windows,'' in \emph{Proceedings of the IEEE/CVF International Conference on Computer Vision (ICCV)}, October 2021, pp. 10\,012--10\,022.

\bibitem{9706678}
A.~Hatamizadeh, Y.~Tang, V.~Nath, D.~Yang, A.~Myronenko, B.~Landman, H.~R. Roth, and D.~Xu, ``Unetr: Transformers for 3d medical image segmentation,'' in \emph{2022 IEEE/CVF Winter Conference on Applications of Computer Vision (WACV)}, 2022, pp. 1748--1758.

\bibitem{Huang_2017_CVPR}
G.~Huang, Z.~Liu, L.~van~der Maaten, and K.~Q. Weinberger, ``Densely connected convolutional networks,'' in \emph{Proceedings of the IEEE Conference on Computer Vision and Pattern Recognition (CVPR)}, July 2017.

\bibitem{shaker2023unetr}
A.~M. Shaker, M.~Maaz, H.~Rasheed, S.~Khan, M.-H. Yang, and F.~S. Khan, ``Unetr++: delving into efficient and accurate 3d medical image segmentation,'' \emph{IEEE Transactions on Medical Imaging}, 2024.

\bibitem{pmlr-v97-zhang19d}
\BIBentryALTinterwordspacing
H.~Zhang, I.~Goodfellow, D.~Metaxas, and A.~Odena, ``Self-attention generative adversarial networks,'' in \emph{Proceedings of the 36th International Conference on Machine Learning}, ser. Proceedings of Machine Learning Research, K.~Chaudhuri and R.~Salakhutdinov, Eds., vol.~97.\hskip 1em plus 0.5em minus 0.4em\relax PMLR, 09--15 Jun 2019, pp. 7354--7363. [Online]. Available: \url{https://proceedings.mlr.press/v97/zhang19d.html}
\BIBentrySTDinterwordspacing

\bibitem{WANG2023102687}
\BIBentryALTinterwordspacing
J.~Wang, Y.~Zheng, J.~Ma, X.~Li, C.~Wang, J.~Gee, H.~Wang, and W.~Huang, ``Information bottleneck-based interpretable multitask network for breast cancer classification and segmentation,'' \emph{Medical Image Analysis}, vol.~83, p. 102687, 2023. [Online]. Available: \url{https://www.sciencedirect.com/science/article/pii/S1361841522003152}
\BIBentrySTDinterwordspacing

\bibitem{doi:10.2214/ajr.184.6.01841774}
\BIBentryALTinterwordspacing
S.~C. Partridge, J.~E. Gibbs, Y.~Lu, L.~J. Esserman, D.~Tripathy, D.~S. Wolverton, H.~S. Rugo, E.~S. Hwang, C.~A. Ewing, and N.~M. Hylton, ``Mri measurements of breast tumor volume predict response to neoadjuvant chemotherapy and recurrence-free survival,'' \emph{American Journal of Roentgenology}, vol. 184, no.~6, pp. 1774--1781, 2005, pMID: 15908529. [Online]. Available: \url{https://doi.org/10.2214/ajr.184.6.01841774}
\BIBentrySTDinterwordspacing

\bibitem{ALDHABYANI2020104863}
\BIBentryALTinterwordspacing
W.~Al-Dhabyani, M.~Gomaa, H.~Khaled, and A.~Fahmy, ``Dataset of breast ultrasound images,'' \emph{Data in Brief}, vol.~28, p. 104863, 2020. [Online]. Available: \url{https://www.sciencedirect.com/science/article/pii/S2352340919312181}
\BIBentrySTDinterwordspacing

\bibitem{Clark2013}
\BIBentryALTinterwordspacing
K.~Clark, B.~Vendt, K.~Smith, J.~Freymann, J.~Kirby, P.~Koppel, S.~Moore, S.~Phillips, D.~Maffitt, M.~Pringle, L.~Tarbox, and F.~Prior, ``The cancer imaging archive (tcia): Maintaining and operating a public information repository,'' \emph{Journal of Digital Imaging}, vol.~26, no.~6, pp. 1045--1057, Dec 2013. [Online]. Available: \url{https://doi.org/10.1007/s10278-013-9622-7}
\BIBentrySTDinterwordspacing

\bibitem{Selvaraju_2017_ICCV}
R.~R. Selvaraju, M.~Cogswell, A.~Das, R.~Vedantam, D.~Parikh, and D.~Batra, ``Grad-cam: Visual explanations from deep networks via gradient-based localization,'' in \emph{Proceedings of the IEEE International Conference on Computer Vision (ICCV)}, Oct 2017.

\bibitem{song2021denoising}
\BIBentryALTinterwordspacing
J.~Song, C.~Meng, and S.~Ermon, ``Denoising diffusion implicit models,'' in \emph{International Conference on Learning Representations}, 2021. [Online]. Available: \url{https://openreview.net/forum?id=St1giarCHLP}
\BIBentrySTDinterwordspacing

\bibitem{NEURIPS2022_260a14ac}
\BIBentryALTinterwordspacing
C.~Lu, Y.~Zhou, F.~Bao, J.~Chen, C.~LI, and J.~Zhu, ``Dpm-solver: A fast ode solver for diffusion probabilistic model sampling in around 10 steps,'' in \emph{Advances in Neural Information Processing Systems}, S.~Koyejo, S.~Mohamed, A.~Agarwal, D.~Belgrave, K.~Cho, and A.~Oh, Eds., vol.~35.\hskip 1em plus 0.5em minus 0.4em\relax Curran Associates, Inc., 2022, pp. 5775--5787. [Online]. Available: \url{https://proceedings.neurips.cc/paper_files/paper/2022/file/260a14acce2a89dad36adc8eefe7c59e-Paper-Conference.pdf}
\BIBentrySTDinterwordspacing

\bibitem{lu2022dpm}
C.~Lu, Y.~Zhou, F.~Bao, J.~Chen, C.~Li, and J.~Zhu, ``Dpm-solver++: Fast solver for guided sampling of diffusion probabilistic models,'' \emph{arXiv preprint arXiv:2211.01095}, 2022.

\bibitem{NEURIPS2023_9c2aa1e4}
\BIBentryALTinterwordspacing
W.~Zhao, L.~Bai, Y.~Rao, J.~Zhou, and J.~Lu, ``Unipc: A unified predictor-corrector framework for fast sampling of diffusion models,'' in \emph{Advances in Neural Information Processing Systems}, A.~Oh, T.~Naumann, A.~Globerson, K.~Saenko, M.~Hardt, and S.~Levine, Eds., vol.~36.\hskip 1em plus 0.5em minus 0.4em\relax Curran Associates, Inc., 2023, pp. 49\,842--49\,869. [Online]. Available: \url{https://proceedings.neurips.cc/paper_files/paper/2023/file/9c2aa1e456ea543997f6927295196381-Paper-Conference.pdf}
\BIBentrySTDinterwordspacing

\bibitem{pmlr-v37-sohl-dickstein15}
\BIBentryALTinterwordspacing
J.~Sohl-Dickstein, E.~Weiss, N.~Maheswaranathan, and S.~Ganguli, ``Deep unsupervised learning using nonequilibrium thermodynamics,'' in \emph{Proceedings of the 32nd International Conference on Machine Learning}, ser. Proceedings of Machine Learning Research, F.~Bach and D.~Blei, Eds., vol.~37.\hskip 1em plus 0.5em minus 0.4em\relax Lille, France: PMLR, 07--09 Jul 2015, pp. 2256--2265. [Online]. Available: \url{https://proceedings.mlr.press/v37/sohl-dickstein15.html}
\BIBentrySTDinterwordspacing

\bibitem{pmlr-v139-nichol21a}
\BIBentryALTinterwordspacing
A.~Q. Nichol and P.~Dhariwal, ``Improved denoising diffusion probabilistic models,'' in \emph{Proceedings of the 38th International Conference on Machine Learning}, ser. Proceedings of Machine Learning Research, M.~Meila and T.~Zhang, Eds., vol. 139.\hskip 1em plus 0.5em minus 0.4em\relax PMLR, 18--24 Jul 2021, pp. 8162--8171. [Online]. Available: \url{https://proceedings.mlr.press/v139/nichol21a.html}
\BIBentrySTDinterwordspacing

\bibitem{song2020score}
Y.~Song, J.~Sohl-Dickstein, D.~P. Kingma, A.~Kumar, S.~Ermon, and B.~Poole, ``Score-based generative modeling through stochastic differential equations,'' \emph{arXiv preprint arXiv:2011.13456}, 2020.

\bibitem{cao2024survey}
H.~Cao, C.~Tan, Z.~Gao, Y.~Xu, G.~Chen, P.-A. Heng, and S.~Z. Li, ``A survey on generative diffusion models,'' \emph{IEEE Transactions on Knowledge and Data Engineering}, 2024.

\bibitem{xu2023restart}
Y.~Xu, M.~Deng, X.~Cheng, Y.~Tian, Z.~Liu, and T.~Jaakkola, ``Restart sampling for improving generative processes,'' \emph{Advances in Neural Information Processing Systems}, vol.~36, pp. 76\,806--76\,838, 2023.

\end{thebibliography}


\vfill
\clearpage


\section*{PGDiffSeg model}

\subsection{Definition of the problem}
The Diffusion Probabilistic Models (DPM) technique utilizes a Markov chain to convert a known distribution into a desired distribution through the forward and reverse processes. Our model is based on the DDPM theory, which can be found in detail in the works of \cite{pmlr-v37-sohl-dickstein15} and \cite{NEURIPS2020_4c5bcfec}.

The intermediate process of our model, illustrated in Fig. \ref{fig1:process_of_diffusion}, entails progressively perturbing the segmentation label until it transforms into a Gaussian distribution, followed by the restoration of the target segmentation
from the Gaussian noise. To precisely identify an image's lesion area and get the desired outcomes, we incorporate the image as a condition into the model, thereby generating a corresponding segmentation map. 

In the forward process, we add Gaussian noise at each timestep, according to a variance schedule $\beta_t$, t=1,..., T, as shown in Equation \ref{eq01}.

\begin{equation}
q\left(x_t \mid x_{t-1}\right):=N\left(x_t ; \sqrt{1-\beta_t} x_{t-1}, \beta_t I\right)
\label{eq01}
\end{equation}
Then distribution in every timestep can be calculated given initial distribution $x_t$, as Equation \ref{eq02}.
\begin{equation}
q\left(x_{1: T} \mid x_0\right):=\prod_{t=1}^T q\left(x_t \mid x_{t-1}\right)
\label{eq02}
\end{equation}
It can be further simplified according to the property of the forward process and get a distribution of arbitrary timesteps more conveniently. With $\alpha_t:=1-\beta_t$ and $\bar{\alpha}_t:=\prod_{s=1}^t \alpha_s$ , it can be defined as Equation \ref{eq03}.
\begin{equation}
q\left(x_t \mid x_0\right)=N\left(x_t ; \sqrt{\bar{\alpha}_t} x_0,\left(1-\bar{\alpha}_t\right) I\right)
\label{eq03}
\end{equation}
Therefore we can skip the calculation of the intermediate distribution $x_{t-1} \ldots, x_1$ to obtain the arbitrary required timestep distribution directly, as Equation \ref{eq04}. For easier understanding, $x_t$ can also be represented as\cite{pmlr-v139-nichol21a} :
\begin{equation}
x_t=\sqrt{\bar{\alpha}_t} x_0+\sqrt{1-\bar{\alpha}_t} \varepsilon \quad \varepsilon \sim N(0, I)
\label{eq04}
\end{equation}
In the reverse process, we need to get $x_{t-1}$ given $x_t$ for every timestep $t$, and this is also what the model needs to learn, as Equation \ref{eq05}.
\begin{equation}
p_\theta\left(x_{t-1} \mid x_t\right):=N\left(x_{t-1} ; \mu_\theta\left(x_t, t\right), \sigma_t^2\right)
\label{eq05}
\end{equation}
Thus every timestep of $x_t$ can be captured, so we can further obtain the target sample, as Equation \ref{eq06}.
\begin{equation}
p_\theta\left(x_0\right):=p\left(x_T\right) \prod_{t=1}^T p_\theta\left(x_{t-1} \mid x_t\right)
\label{eq06}
\end{equation}

The variance here is still learnable. The relationship between the two extreme choices of variance $\beta_t$ and $\tilde{\beta}_{t} = \frac{1-\bar{\alpha}_{t-1}^{-}}{1-\bar{\alpha}_{t}} \beta_{t}$ had similar results. \cite{NEURIPS2020_4c5bcfec} also indicates that it is better to use a fixed variance because learning reverse process variance can lead to instability in the training process and poor sample quality. So we refer to the DDPM setting that the variation of $\beta_t$ can be linear, from $\beta_1$=1×10-4 to $\beta_T$=2×10-2. Following \cite{NEURIPS2020_4c5bcfec}, we can also parameterize $\varepsilon_\theta$ instead of $\mu_t$ according to Equation \ref{eq07}.
\begin{equation}
\mu_\theta\left(x_t, t\right)=\frac{1}{\sqrt{\alpha_t}}\left(x_t-\frac{\beta_t}{\sqrt{1-\bar{\alpha}_t}} \varepsilon_\theta\left(x_t, t\right)\right)
\label{eq07}
\end{equation}
and the sample $x_{t-1} \sim p_\theta\left(x_{t-1} \mid x_t\right)$ can be expressed as Equation \ref{eq08}.

\begin{equation}
\begin{aligned}
x_{t-1} =& \frac{1}{\sqrt{\alpha_t}}\left(x_t-\frac{\beta_t}{\sqrt{1-\bar{\alpha}_t}} \varepsilon_\theta\left(x_t, t\right)\right) + \sigma_t z, \\
& \text{where} \quad z \sim N(0, I)
\end{aligned}
\label{eq08}
\end{equation}

Our choice is consistent with DDPM that predicts the noise $\varepsilon$ to produce $\mu_\theta\left(x_t, t\right)$ except for using MSE loss to optimize the model parameters as Equation \ref{eq09}.

\begin{equation}
\operatorname{loss}(\mathrm{x}, \mathrm{y})=\frac{1}{\mathrm{n}} \sum\left(\mathrm{x}_{\mathrm{i}}-\mathrm{y}_{\mathrm{i}}\right)^2
\label{eq09}
\end{equation}

\begin{figure}[htbp]
\centering
\includegraphics[scale=.5]{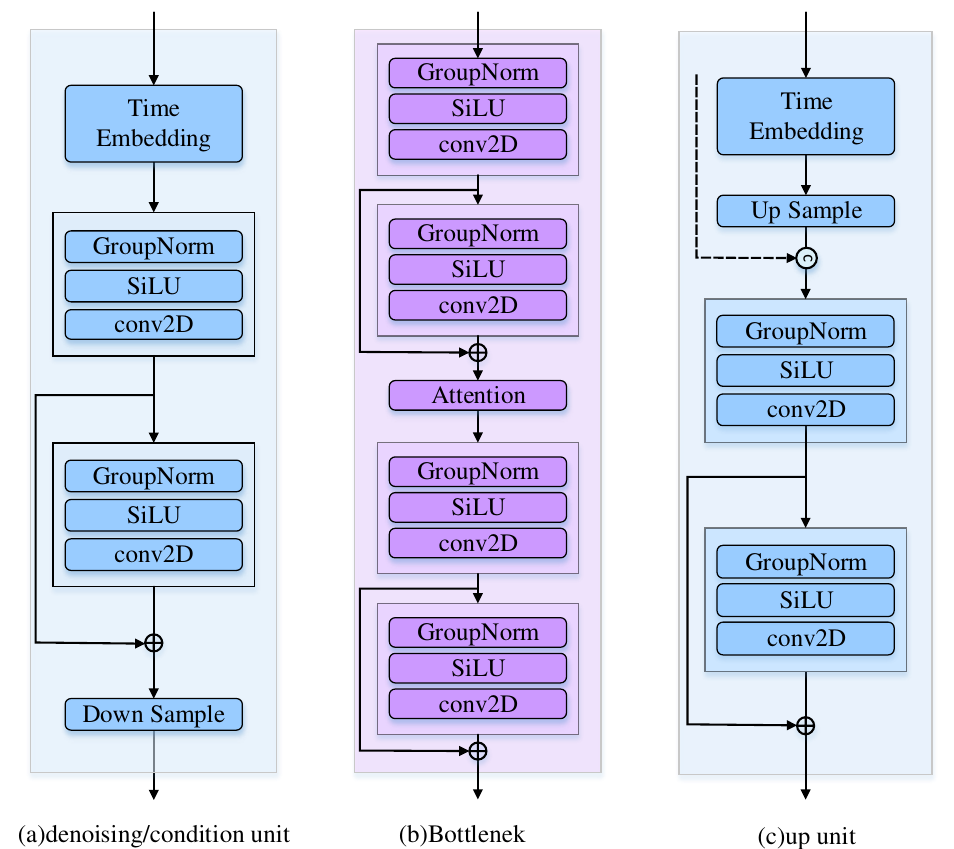}
\caption{Three sub-modules in the PGDiffSeg model. "(a) denoising/condition unit" is the details of the module for denoising flow and condition flow in the feature encoder module, as shown in Fig. \ref{fig2:Overall-view-of-our-model}(b). The denoising unit and condition unit have the same structure.
"(b) Bottlenek" is a detailed presentation of Fig. \ref{fig2:Overall-view-of-our-model}(d). "(c) Up unit" is the implementation of the component in Fig. \ref{fig2:Overall-view-of-our-model}(e).}
\label{fig3:model-part}
\end{figure}

\subsection{Denoising and condition units}
\label{3.4.1}
The condition flow uses the extracted image features to transmit spatial features, while the denoising flow uses the extracted $x_t$ features to transmit denoised features. Due to the similar spatial structure between the image and the noisy label, we designed the same unit for encoding both the image and the noised label, as shown in Fig. \ref{fig3:model-part}(a). Inspired by \cite{NEURIPS2020_4c5bcfec} and \footnote{\url{https://github.com/abarankab/DDPM}}, we added time embedding to capture the temporal order, allowing multiple time steps of the diffusion process to share the same set of model parameters. 

The denoising and condition units include time embedding, residual blocks, and downsampling. The residual block is composed of two blocks with a residual connection. The definition of the block is as Equation \ref{eq102}. 

\begin{equation}
H(\cdot)=\operatorname{GroupNorm}(\operatorname{SiLU}(\operatorname{Conv}(\cdot)))
\label{eq102}
\end{equation}

As shown on the left side of Fig. \ref{fig2:Overall-view-of-our-model}, the denoising flow performs convolution with $x_t$ to accomplish denoising. In contrast, the condition flow performs convolution with the image and holds crucial information for denoising, significantly impacting the ultimate denoising outcome. Conversely, the denoising flow also carries tumor-related information that can emphasize the tumor's location in the condition flow.

\subsection{Bottleneck module}
The Bottleneck module receives and fuses features from the feature encoding and prior knowledge-guided modules. This fusion is performed to inject prior tumor information, which guides the learning process of the segmentation task.

The Bottleneck is a vital component in connecting the encoder and decoder, making it indispensable for the model. As shown in Fig. \ref{fig3:model-part}(b), the Bottleneck comprises two residual blocks, as described in Section \ref{3.4.1}. Additionally, we introduce a self-attention module between these two residual blocks. This inclusion enables improved integration of high-level semantic features from the encoder and provides the decoder with richer contextual information.

\subsection{Feature decoder}
\label{3.7}
Feature decoder plays a critical role in reconstructing tumor details. It converts the high-level feature representations extracted from the Bottleneck into the output corresponding to the original $x_t$. By leveraging skip connections, it integrates the abstract features extracted by the encoder to restore the details and structure of the initial input gradually. In this part, we refer to the previous work and design the Feature decoding using four up units and a convolutional block. The structure of the up unit is described in Fig. \ref{fig3:model-part}(c). It first encodes the input with timesteps $T$, then increases the size of the feature map through upsampling. Next, a skip connection is used with the outputs of the corresponding units in feature encoding. Finally, it undergoes feature decoding using the residual blocks defined in the Denoising and Condition Units section. After all the up units, we restore the channel dimensions of the feature map to the original size through convolutional operations.

\section*{Efficient sampling}
According to \cite{song2020score}, the noise perturbations used in our diffusion model framework (DDPM) can be viewed as discretizations of a Stochastic Differential Equation (SDE). When DDPM is continuous, it can be formulated as a Variance-Preserving SDE, meaning that as the number of steps approaches infinity, its variance remains bounded. Additionally, \cite{song2020score} demonstrates that for all diffusion processes, a deterministic process exists whose trajectories have the same marginal probability density as the SDE. Consequently, the SDE used for modeling also has an equivalent Ordinary Differential Equation (ODE) form, known as the probability flow ODE.

Using this equivalent ODE form during sampling typically results in higher sampling efficiency and allows for precise likelihood computation of the samples generated by the model. Compared to SDEs, probability flow ODEs can be solved with larger step sizes because they have no randomness \cite{cao2024survey}. Consequently, many works have leveraged advanced ODE solvers \cite{song2021denoising, NEURIPS2022_260a14ac, lu2022dpm, NEURIPS2023_9c2aa1e4} to achieve faster sampling speeds. These methods can reduce the number of function evaluations by over 90$\%$ compared to the original DDPM sampler, significantly speeding up the sampling process while still producing high-quality samples. Since ODE samplers introduce less discretization error than SDE samplers  \cite{song2020score, xu2023restart}, most previous work on accelerating sampling has concentrated on ODEs. While ODE-based samplers are generally faster, they have reached their performance limits. On the other hand, SDE-based samplers can provide better sample quality, albeit at slower speeds  \cite{cao2024survey}.

\section*{Details of datasets}
We present some details of the dataset used, including the modalities, image size, grayscale range, the total number of images, and division of training, validation, and test sets for each dataset, as summarized in Table\ref{details-of-two-datasets}.

\begin{table*}
\caption{Details of the two datasets used in this experiment. Note: the Breast-MRI-NACT-Pilot (NACT for brief) dataset consists of three-dimensional data, and we slice it into 2D slices to fit our model's input.}
\label{details-of-two-datasets}
\makebox[\textwidth]{
\begin{tabular}{ccccccc}
\hline
Dataset & Modality       & Image size (w$\times$h) & Grayscale range & Total number/slices & \begin{tabular}[c]{@{}l@{}}Number/slices \\ (with tumor, no tumor)\end{tabular} & \begin{tabular}[c]{@{}l@{}}Number/slices \\ (train, val, test)\end{tabular} \\
\hline
NACT    & MRI, 2D        & 256$\times$256, 512$\times$512 & 0-5196           & 3834         & 2873/961                      & (2634,420,780)            \\
BUSI    & Ultrasound, 2D & Different sizes         & 0-255            & 780          & 647/133                       & (545,78,157)    \\
\toprule[1pt]
\end{tabular}
}
\end{table*}

\section*{Sampling process}
\begin{figure}[htbp]
\centering
\includegraphics[scale=.45, angle=-90]{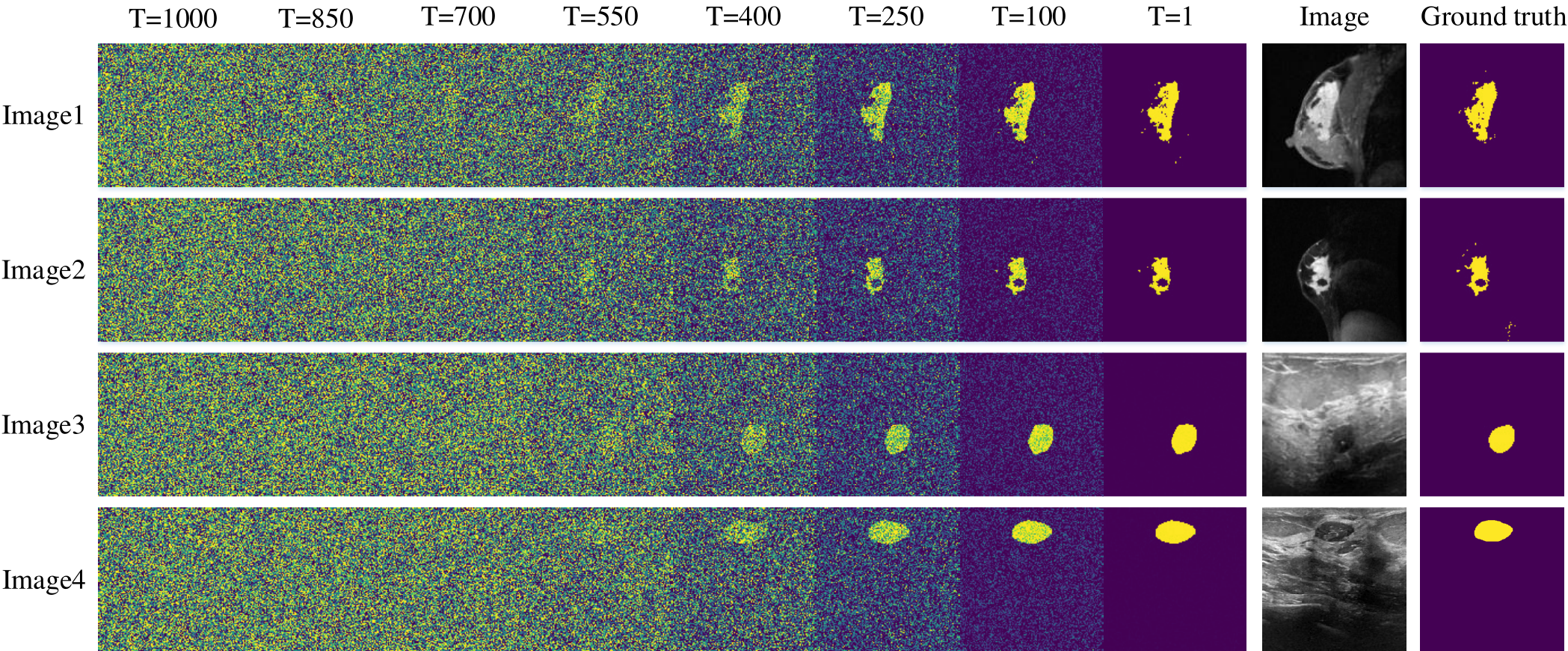}
\caption{Sampling Process of the GFDiffSeg Model, visualized by viridis color mapping between 0 in deep purple and 1 in bright yellow. The trained model starts sampling from Gaussian noise, and the corresponding labels gradually emerge from the Gaussian noise under different images as conditions. We selected two representative images from each dataset. The first two rows are from the Breast-MRI-NACT-Pilot dataset, and the last two are from the BUSI dataset.}
\label{sampling-process}
\end{figure}

The sampling process of the proposed GFDiffSeg model is illustrated in Fig\ref{sampling-process}. It can be observed that the target region gradually emerges from a noise-dominated distribution, indicating the model's capability to iteratively denoise from a state of pure Gaussian noise and accurately reconstruct the segmentation results.

\end{document}